\pdfoutput=1 
\documentclass[english,nohyper,a4paper]{tufte-handout}
\usepackage{amsmath,amssymb,amsthm}

\usepackage[latin9]{inputenc}
\usepackage[T1]{fontenc}
\usepackage{babel}

\usepackage[semibold]{sourcesanspro}

\usepackage{color}
\usepackage{wasysym}   
\usepackage{marvosym}  
\usepackage{graphicx}
\usepackage{microtype}
\usepackage{booktabs}

\def\adx#1:#2\par{\par\halign{\hskip #1##\hfill\cr #2}\par}

\def\mesa{\texttt{MESA}}

\def\teff{T_{\rm eff}}
\def\mast{M_\ast}

\def\msol{M_\odot}

\def\sigr{\sigma_{\rm R}}
\def\sigi{\sigma_{\rm I}}

\def\diff{{\mathrm d}}

\def\unity{ \hbox{1\kern-.23em l} }
\def\zero{ \hbox{0\kern-.23em |} }
\def\field{ \hbox{I\kern-.23em K} }

\def\braket #1.#2.{\langle #1 \vert #2 \rangle}

\def\imag{{\rm i}}

\newcommand{\lyxaddress}[1]{ \vspace{1.4em} \par {\raggedright #1 \vspace{1.4em}
\noindent\par} }

\usepackage{amsmath}
\title{\`A Propos Crossing the Hertzsprung Gap} 
\author{Alfred Gautschy}
\date{}

\begin{document} \maketitle  

\lyxaddress{CBmA, 4410 Liestal, Switzerland}

\begin{abstract} \noindent The evolution of intermediate-mass and massive
stars speeds up considerably after they finish their hydrogen core-burning. 
Due to this accelerated evolution, the probability to observe 
stars during this episode is small. In suitable stellar aggregates, 
in particular star clusters of appropriate ages, the fast evolutionary 
phase causes a depopulated area~--~referred to as the Hertzsprung 
gap~--~in color-magnitude diagrams and derivatives therefrom. 
The explanation of the speed-up usually resorts to the star's 
Kelvin-Helmholtz timescale and the Sch\"onberg-Chandrasekhar 
instability is called upon. This exposition challenges this viewpoint
with counterexamples and argues that a suitably defined nuclear timescale
is enough to explain the fast evolution. A thermal instability, 
even though it develops in stars evolving through the Hertzsprung gap, 
is not a necessary condition to trigger the phenomenon. 
\end{abstract}

\section{Introduction}
\begin{marginfigure} 
	\begin{center}
	\includegraphics[width=1.1\textwidth]{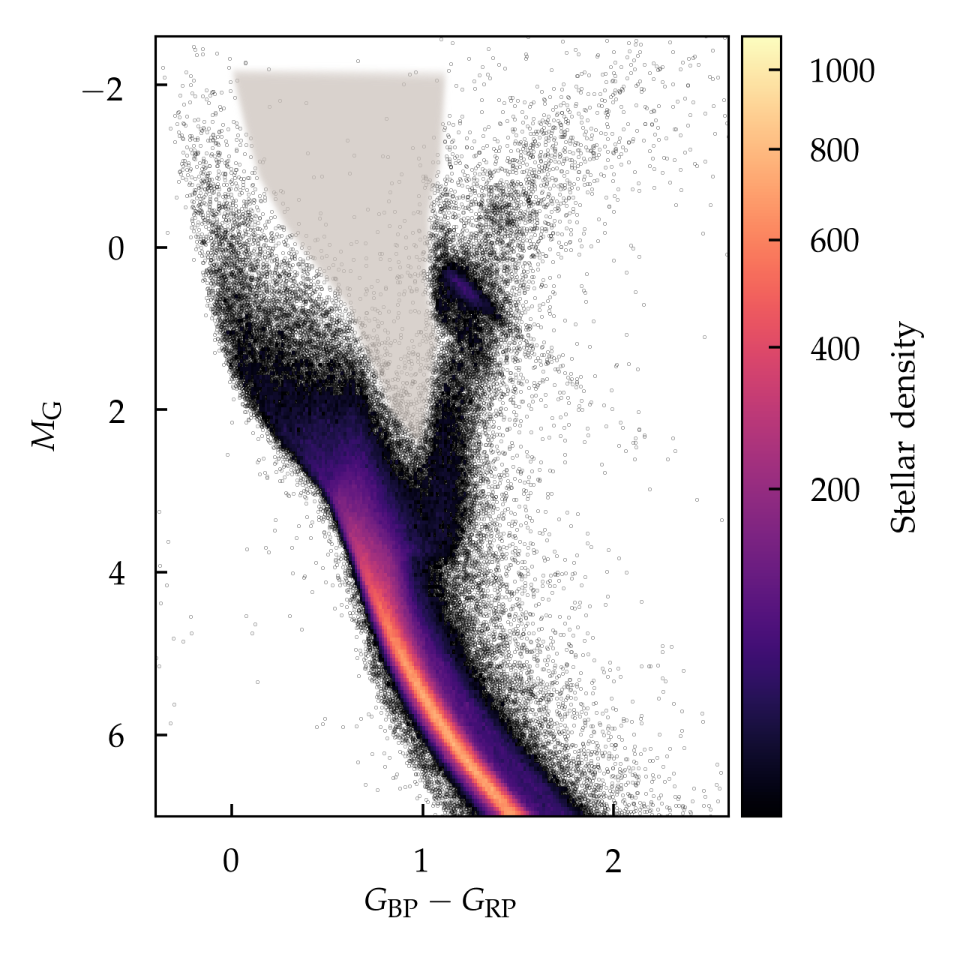} 
	\end{center}
     \caption{Color-magnitude diagram from the stars within
     		  250~pc around the sun as observed with \emph{Gaia }and 
     		  made available in DR2. The grey wedge in the 
     		  upper left hints at the region referred to as the 
     		  \emph{Hertzsprung gap}.
             } \label{fig:GaiaCMD}
\end{marginfigure}

Color-magnitude diagrams~--~and therefore also the Hertzsprung-Russell (HR)
diagrams resorted to in the following~--~of suitably chosen stellar aggregates 
often show a region, originally specified as \emph{between early A-type 
and late G-type stars, which is only scarcely populated.} 
The laudatio of \citet{Phillips1929}, on the occasion of awarding the 
Gold Medal of the Royal Astronomical Society to Hertzsprung, reviewed 
concisely the pertinent work of awardee to
establish the reality of a \emph{gap} in the distribution of the stars
as a function of spectral type. It was the paper of 
\citet{Stroemberg1932} through which the expression `Hertzsprung gap' 
found its way to the astronomical literature. 
Figure~\ref{fig:GaiaCMD} illustrates the phenomenon using a modern 
mixed color-magnitude/Hess diagram comprised of the brighter stars 
within 250~pc of the sun that were measured with the 
astrometric space mission \emph{Gaia}. The observed Hertzsprung gap is 
marked with the grey wedge added to the upper left of the diagram.

In the early 1950s when the general direction of a star's evolution became 
clear \citep[cf.][]{Sandage1952}, it was realized that the Hertzsprung 
gap is a region through which sufficiently massive stars evolve 
comparatively fast after hydrogen is exhausted in their cores. 
The higher evolutionary speed lowers hence the probability to catch 
stars during that episode. Already early on, the origin of the speed-up 
was conjectured to be connected with the
Sch\"onberg-Chandrasekhar (SC) instability.  

At the end of the main-sequence phase, i.e. when the stellar cores run out 
of nuclear fuel, the stars adjust their structure as the nuclear burning 
shifts to a shell on top of a nuclear inactive core. 
At first, the shell is fat but it grows ever thinner as the star 
approaches the base of the first giant branch (FGB). 
In the absence of nuclear burning no luminosity is generated and hence 
the temperature gradient in the core vanishes also~--~at least in the 
clinically clean case of a star in full equilibrium. If the core contracts, 
on the other hand, a small temperature gradient is maintained. 
In particular for intermediate-mass and massive stars, the
core must contract once it grows too massive. If the relative core mass, $q$,
exceeds roughly $0.1$, the SC-instability \citep[e.g.][]{kww12} 
predicts the shrinking and an accompanying heating of the inert stellar core.
The SC thermal instability is usually linked to the Kelvin-Helmholtz timescale
of the
star.\sidenote{\citet{Gabriel1967} were the first to do a full secular stability 
analysis of stellar structures that  correspond to the original SC model setup.
\citet{ledoux69} felt convinced that the fast evolution connecting the terminal
main-sequence and the early FGB phase was due to the identified secular
instability. }
Even if not stated \emph{expressis verbis}, the
students tend come away from the textbooks with the conceptual thread
`Hertzsprung gap'~--~`Kelvin-Helmholtz timescale'~--~`SC-instability'. 
But is this really a causal chain? This question has been 
contemplated and debated repeatedly in the past 
\citep[e.g.][and references therein] {LauterbornSiquig76, 
SpindlerLauterborn1977, Hansen1978}. 
The answers pointed mostly to a circumstantial connection of 
the SC instability and the evolutionary speed-up of the stars
across the HR plane.

The following exposition presents the results of counterexamples to the
reasoning of thermal instability being the cause of the speed-up.  
Even though hardly anything is new, the discussion and
the compilation of facts put forth should help students 
of the stars to shape their understanding of the phenomenon and hopefully 
allows them to proceed beyond where we currently stall. 
An appendix is devoted to the presentation 
of canonical secular stability computations, which are referred to in the
main text, addressing their usefulness 
and their limitations. In contrast to the older literature, the acquired 
stability results are comprehensive enough to illustrate
how the lowest-order secular modes evolve and interact.

\section{Stellar Evolution from the ZAMS to the FGB}

\begin{figure} 
	\begin{center}
	\includegraphics[width=0.8\textwidth]{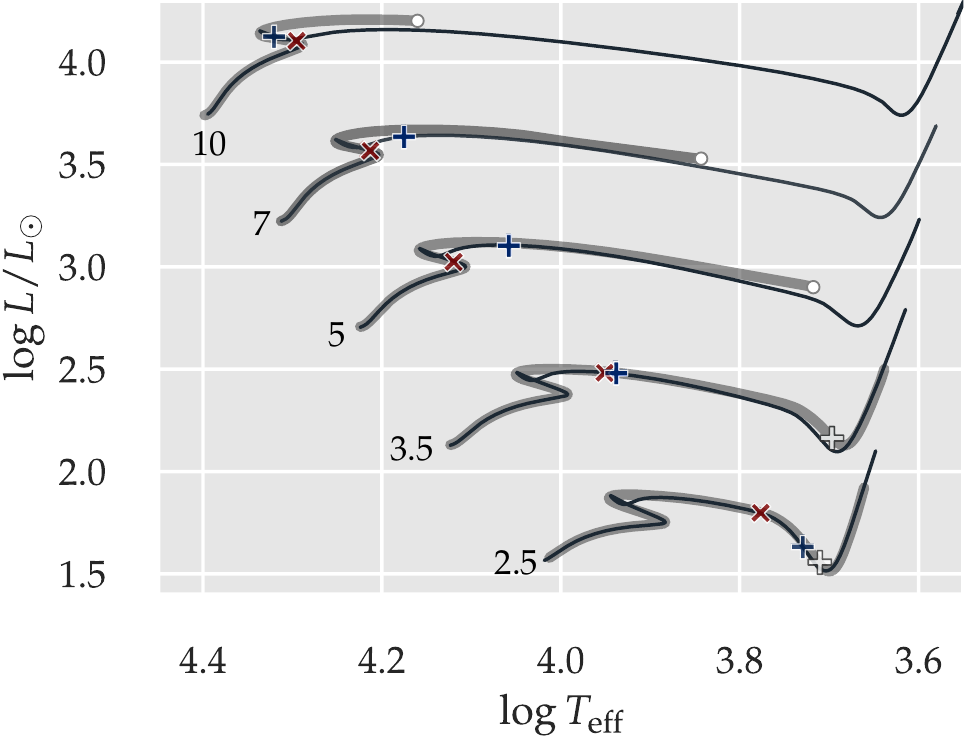} 
	\end{center}
     \caption{A collection of evolutionary tracks of intermediate-mass  and
     		  massive stars in QHE (thin back) and FE (thick grey
	          lines) approximation. The stars' masses in solar units
     		  are labeled along the ZAMS. The meaning of the crosses and
     		  the white circles is elaborated in the text.
             } \label{fi:AllTracks}
\end{figure}

The evolutionary tracks on the HR Diagram of the intermediate-mass 
and massive model stars ($2.5 < \mast/\msol < 10$) that were studied 
for this note are shown in Fig.~\ref{fi:AllTracks}. The thin black lines 
trace the \mesa-computed quasi-hydrostatic evolution (QHE). 
The QHE computations were arbitrarily stopped along the 
FGB at central helium ignition. The stars' initial homogeneous composition 
was set to $(X,Z) = (0.7, 0.02)$. Boundaries of the convection zones were 
determined using Ledoux's 
criterion.\marginnote[-1.5cm]{Further choices in the \texttt{\&controls} 
			inlist were: \texttt{mixing\_length\_alpha=1.8}
            and for some mild overshooting:
            \texttt{overshoot\_f$\langle \ast \rangle$=0.012}
            \texttt{overshoot\_f0$\langle \ast \rangle$=0.002}
            with the wildcard $\langle \ast \rangle$ covering 
            all necessary subspecifications.
		    The particular choices to model the stars change the 
		    numerical values of the results presented here but 
		    not their qualitative behavior. 
           }

Additionally, the evolution of stars in full equilibrium (FE), 
i.e. imposing $\diff_t Q \equiv 0$ in the energy equation, 
was calculated for the same set of stellar masses (heavy grey lines in Fig.~\ref{fi:AllTracks}). 
The computation of two lowest-mass FE tracks was stopped 
arbitrarily once the stars started to ascend the FGB. 
The \emph{white crosses }along the lower-mass FE model evolutionary tracks
mark the epochs when a monotonic secular
instability\sidenote[][-3mm]{computed with the linear perturbation analysis 
                     discussed in the Appendix}
sets in. Evidently, the computation of these two lower-mass FE model 
sequences was not impeded by the instability of the lowest-order secular mode.
The higher-mass models ($\mast > 3.5\,\msol$), on the other hand, 
stalled~--~indicated by a 
\emph{white circle }at the end of the respective tracks~--~when the Henyey-type 
solver failed to converge. In FE modeling, this computational breakdown 
is attributed to the onset of a monotonic thermal instability of 
the model once the uniqueness of the equilibrium solution is violated 
\citep[cf.][]{Gabriel1967, Paczynski1972a}.
At least for sufficiently low stellar masses, 
\mesa~in FE-mode is apparently able to proceed past the onset of 
a monotonic thermal instability. In \mesa, the stellar
structure equations are solved \emph{simultaneously} with those of
the nuclear network and the mixing of nuclear 
species \citep[see Fig.47 in][]{Paxton2013}.
Therefore, depending on the number of explicitly treated 
species,\sidenote{The computations in this study used the \texttt{approx21.net}
	              network. The compositional subspace entering the Henyey matrix,	
	              $$
	              \frac{\mathrm{d}(\mathrm{chem}_k)}{\mathrm{d}(\mathrm{chem}_k)} 
	              \in M(21 \times 21)\,,
	              $$ in \citet{Paxton2013} notation, is considerably larger
	              than the structural $M(4 \times 4)$ one. The subscript
	              refers to cell $k$ of the spatial discretization.
	             }
the size of a Henyey block can be considerably larger than in the traditional
stellar evolution codes. Even if the determinant of the `structure subspace'~--~matrix 
vanishes once a real secular eigenvalues passes through zero, 
the determinant of the total Henyey matrix may, nonetheless, 
remain nonzero due to the effect of the compositional subspace. 
The detailed reason of why lower-mass FE model stars avoid stalling
at the onset of the monotonic secular instability after their main-sequence phase
but the more massive siblings fail remains to be understood.

The \emph{red crosses} ($\times$) in Fig.~\ref{fi:AllTracks} mark the epochs 
when the relative helium core-mass exceeds $q_{\mathrm{core}}= 0.1$. 
This choice of $q_{\mathrm{core}}$ was motivated by the magnitude usually quoted 
when ideal-gas model envelopes fail to match isothermal cores, 
i.e. when the SC instability is encountered in respective simple model stars. 
The $5$ to $10\,\msol$ sequences shown in Fig.~\ref{fi:AllTracks} start 
their hydrogen-free core evolution with a helium core whose mass already exceeds
$q_{\mathrm{core}}= 0.1$. The red crosses mark therefore the epochs of 
the emergence of inert pure helium cores that are initially already
SC super-critical. 

Finally, the \emph{ blue crosses} ($+$) mark the onset 
of the monotonic secular instability as obtained from linear 
secular stability computations (see Appendix) performed on QHE models, 
this is the instability  that is 
usually associated with the SC instability \citep[e.g.][]{Kaehler1972}. 
The secular eigenanalyses covered the evolutionary stages from the ZAMS 
to the lower FGB. Red edges of the unstable monotonic modes were not encountered in
any of the sequences.

\begin{marginfigure} 
	\begin{center}
	\includegraphics[width=1.1\textwidth]{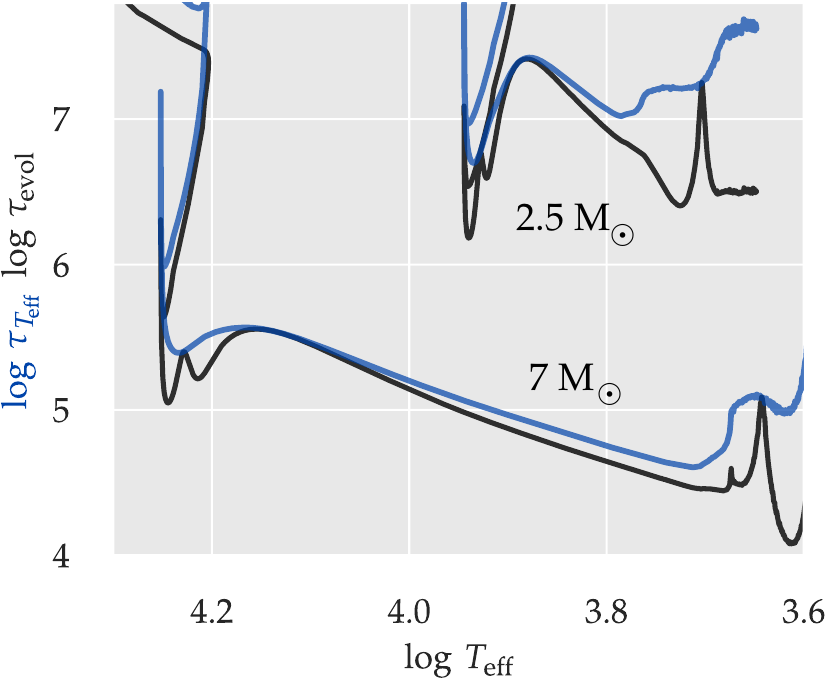} 
	\end{center}
     \caption{ 
     Variation and comparison of characteristic timescales during the
     post main-sequence evolution. The transverse timescale,
     $\tau_{\teff}$~(blue), is the upper bound to the evolutionary 
     timescale, $\tau_{\rm{evol}}$ (black), which accounts also for the 
     luminosity change.
     The black peaks touching the $\tau_{\teff}$ loci at low $\teff$ 
     mark the base of the FGB when the stars' luminosity passes through  
     a local minimum.
             }  \label{fig:Cmpre_Speed}
\end{marginfigure}

To quantify the depopulation of stars along the effective-temperature axis 
we introduce the \emph{transverse evolution }timescale of stars across 
the HR diagram as: 
\begin{equation*}
 \tau_{\teff} = \frac{\Delta t}{\left|\Delta \ln\teff\right|}\,. 
\end{equation*}
\marginnote[-1cm]{for some suitably chosen time interval $\Delta t$ 
                  and the accompanied $\Delta \teff(\Delta t)$}
The quantity $\tau_{\teff}$ measures how efficiently stars get plowed out of
selected temperature/color bins.
If a star evolves mostly horizontally in the HR diagram 
then $\tau_{\teff}$ is even an indicator of the overall evolution timescale on 
that plane. 
\begin{equation*}
 \tau_{\mathrm{evol}} = \frac{\Delta t}{\left(\left|\Delta \ln L\right|^2 + \left|\Delta \ln\teff\right|^2 \right)^{1/2}}
 \,. 
\end{equation*}
On the other hand, if also a star's luminosity changes significantly,
$\tau_{\teff}$ still serves as an upper bound of the evolution timescale. 
This latter case is encountered at low effective temperatures of lower-mass stars
where $\tau_{\teff}$ first remains high because it is blind to the luminosity
drop before the base of the FGB is reached. For higher stellar masses this
discrepancy shrinks as illustrated in Fig.~\ref{fig:Cmpre_Speed}.

\begin{figure} 
	\begin{center}
	\includegraphics[width=0.75\textwidth]{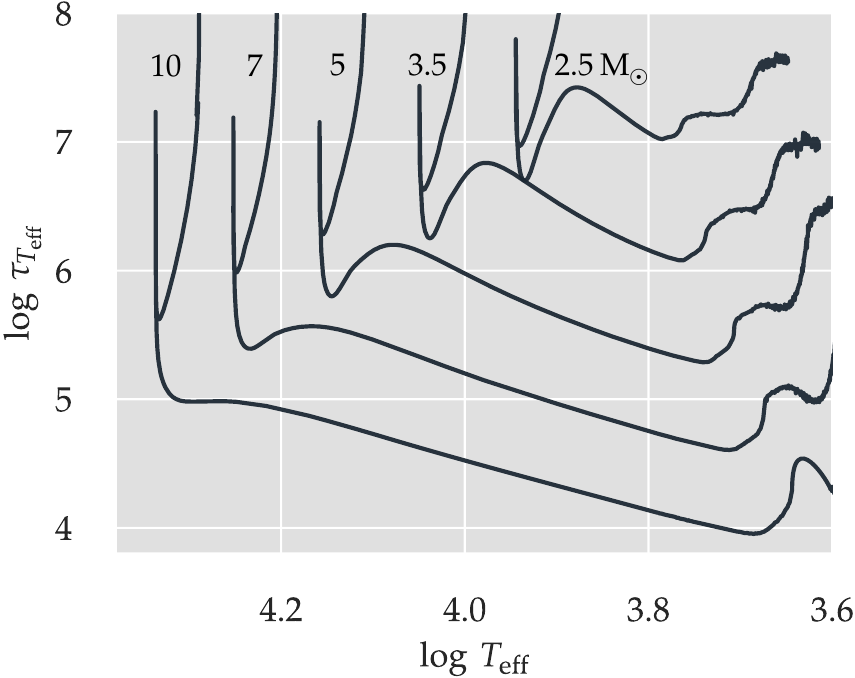} 
	\end{center}
     \caption{The variation of the timescale $\tau_{\teff}$, a measure of
     the magnitude of transverse evolution across the HR diagram, 
     for the QHE model sequences presented in Fig.~\ref{fi:AllTracks}. 
     After their first turn of the S-bend evolution the model
     stars enter the diagram at the top. For all models more massive 
     than about $2.5\,\msol$, $\tau_{\teff}$ drops significantly, 
     i.e. evolution speeds up, when they develop 
     their H-shell burning and evolve to the base of the FGB.    
             }  \label{fig:EvolSpeed}
\end{figure}

The $\tau_{\teff}$ behavior associated with the QHE evolutionary tracks displayed 
in Fig.~\ref{fi:AllTracks} is collected in Fig.~\ref{fig:EvolSpeed}. 
The stars enter the diagram on the top left of the respective loci after 
they have taken the first turn of their S-bend evolution; i.e. when the stars' 
central furnace runs out of hydrogen. The first speedy evolution episode 
develops while stars evolve through the second turn of the S-bend 
(the bowl-shaped episode in Fig.~\ref{fig:EvolSpeed}) while the 
hydrogen-burning shell develops. 
For stars more massive than about $2.5\,\msol$ the post~--~S-bend evolution 
to lower effective temperatures gets progressively faster towards the base 
of the FGB where $\tau_{\teff}$ passes through a local minimum. For the set 
of computed evolutionary tracks the margin table reports the maximum speed-up 
of the stars' evolution relative to the epoch around the end of their S-bend phase. 
Notice that the speed-up by a factor two in the $2.5\,\msol$ model sequence is 
reached already shortly after the end of the S-bend passage. 
The subsequent evolution to the base of the FGB is mostly slower than 
that through the S-bend. Together with the inspection of Fig.~\ref{fig:EvolSpeed} 
we deduce that~--~for the set of parameters chosen for this study~--~the
Hertzsprung-gap develops for stars more massive than about $2.5\,\msol$. 
After $\tau_{\teff}$ evolved through its local minimum at low effective temperatures 
it eventually forfeits its usefulness because the steep luminosity rise 
dominates the stars' evolution along the FGB (cf. Fig.\ref{fig:Cmpre_Speed}).
\begin{margintable}[-4cm]
  \begin{tabular}{c c}
	\toprule
	Mass/$\msol$&  max. transverse speed-up \\
	\midrule
	 2.5 & 2   \\
	 3.5 & 3.5 \\
	 5   & 10  \\
	 7   & 25  \\
	10   & 40  \\
	\bottomrule
  \end{tabular} \label{tbl:SpeedUp}
\end{margintable}

Evidently, Fig.~\ref{fig:EvolSpeed} and the rough quantification in the 
margin-table illustrate how QHE modeling accounts for the relative scarcity of
stars in the Hertzsprung gap of the HR diagrams. The QHE models 
predict qualitatively that for stellar masses $\gtrsim 2.5 \msol$ depopulation 
starts close to the main sequence band. With increasing mass, 
at higher luminosity hence, the gap broadens. The lowest $\tau_{\teff}$ 
is encountered close to the base of the FGB.

\begin{figure} 
	\begin{center}
	\includegraphics[width=0.75\textwidth]{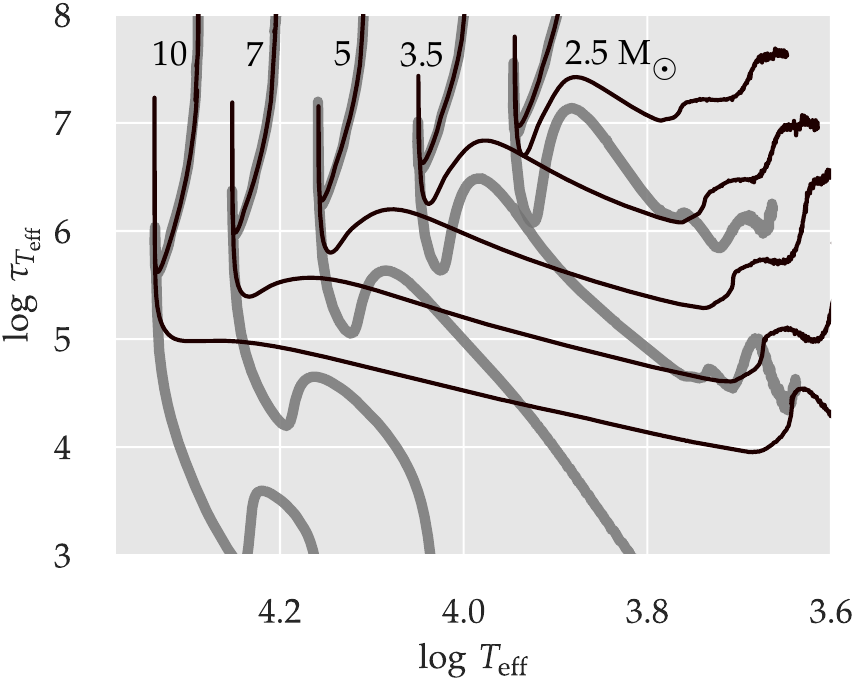} 
	\end{center}
     \caption{ Comparing the QHE-models' $\tau_{\teff}$ (thin black lines)
     	       with those of the respective FE-models (thick grey lines) 
     	       reveals the latter to evolve even faster across 
     	       the HR plane than their QHE brethren.                
             }  \label{fig:EvolSpeed_QHEandFE}
\end{figure}

The QHE tracks and the associated transverse evolutionary timescales account 
the Hertzsprung gap. If the origin of the accelerated evolution after 
central H-burning is indeed caused by a thermal instability (which is 
present in the QHE ansatz via the $\diff_t Q$ term in the energy equation) 
then post~--~main-sequence evolution should slow down in the FE approximation.
Thanks to \mesa 's property to evolve model stars in FE mode, we could
follow lower-mass stars 
($\mast \lesssim 3.5 \msol$) even up the lower FGB and the more massive model sequences
at least to the onset of their secular instability. Therefore, \mesa~offers the unique
opportunity to compare FE $\tau_{\teff}$ values to the respective QHE numbers 
over a pertinent range of evolutionary stages. 
Figure~\ref{fig:EvolSpeed_QHEandFE} shows again the set of $\tau_{\teff}$ 
loci of the QHE evolution as thin black lines and those from FE evolution 
superimposed as heavy grey lines. 
The correspondence of QHE and FE loci are evident because 
they are essentially identical to the end of the S-bend evolution. 
Once the H-burning shell developed, FE-evolution across 
the HR plane is, for all masses considered, \emph{always }faster than 
the corresponding QHE  
one.\sidenote{This was pointed out already by \citet{Roth1973}} 
Since $\diff_t Q = 0$ underlies the FE models, 
the fast FE evolution cannot be attributed to thermal imbalance 
and hence is \emph{not }caused by a thermal instability. Apparently, thermal 
imbalance slows down post~--~main-sequence evolution as
compared to models in thermal equilibrium.  

\medskip

\begin{figure*}
\label{fig:Timescales}
\centering
\includegraphics[width=0.46\linewidth]{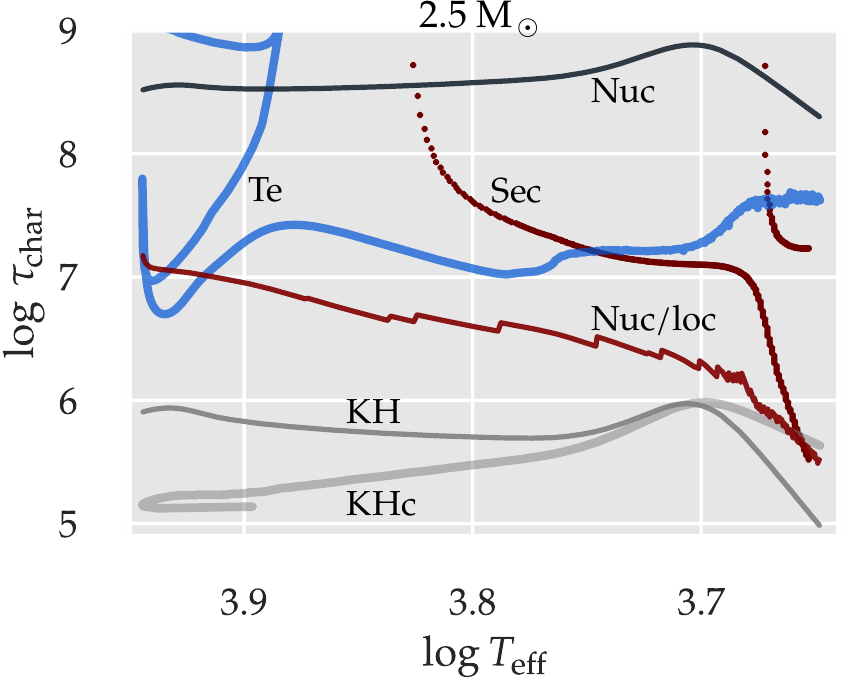}
\hspace{0.3cm}
\includegraphics[width=0.46\linewidth]{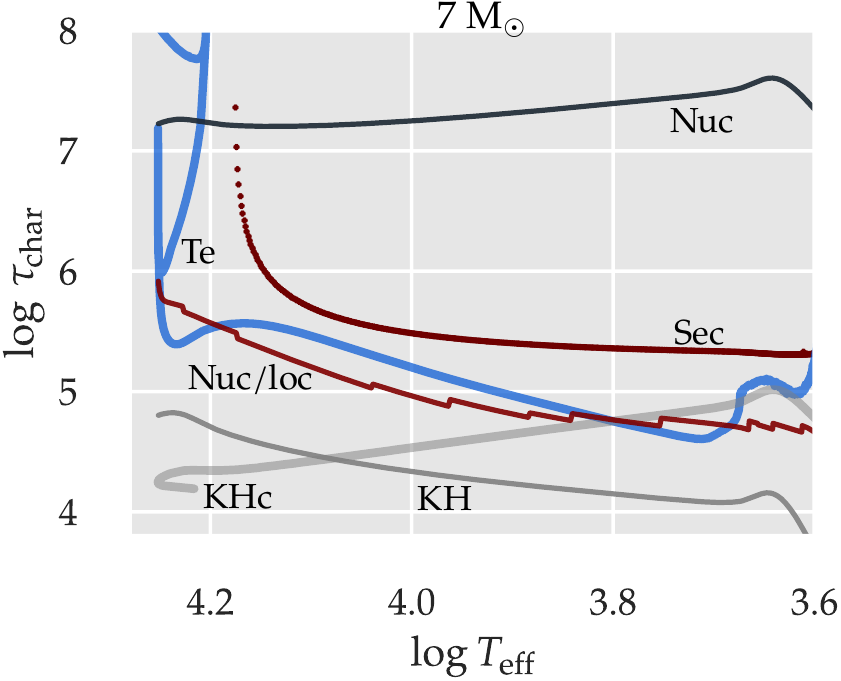}
\caption{Comparison of various \emph{timescale} concepts
along two representative evolutionary paths. The left
panel shows the case of the $2.5 \msol$ star at the low-mass boundary of 
the Hertzsprung gap; the right panel that of the
$7 \msol$ one. The meaning of nuclear (Nuc), local nuclear (Nuc/loc), 
Kelvin-Helmholtz (KH), secular (Sec), and the transverse evolution 
timescales are discussed in the text.}
\end{figure*}

Even though we tried to get away from correlations, the transverse evolutionary
timescales be compared in the following to some  
timescales canonically resorted to in stellar astrophysics: 
Figure~\ref{fig:Timescales} compares for the representative $2.5\,\msol$ and 
$7\,\msol$ tracks the models' nuclear timescale, $\tau_{\mathrm{Nuc}}$ (Nuc), 
a suitably defined \emph{local }nuclear timescale, $\tau_{\mathrm{Nuc/loc}}$ 
(Nuc/loc), the Kelvin-Helmholtz timescale, $\tau_{\mathrm{KH}}$ (KH), 
of the complete stars, and the Kelvin-Helmholtz timescale of the cores 
only, $\tau_{\mathrm{KHc}}$ (KHc): 

\begin{align*}
 \tau_{\mathrm{Nuc}}  & = 1\cdot 10^{10} \frac{M_\ast/M_\odot}{L_\ast/L_\odot}\,, \\
 \tau_{\mathrm{Nuc/loc}} & = \left|\frac{X}{\dot{X}}\right|_{\varepsilon_\mathrm{max}}\,,  \\
 \tau_{\mathrm{KH}}   & = 3 \cdot 10^7 \frac{(M_\ast/M_\odot)^2}
                                            {(R_\ast/R_\odot)(L_\ast/L_\odot)}\,, \\
 \tau_{\mathrm{KHc}}  & = 3 \cdot 10^7 \frac{(M_{\mathrm{core}}/M_\odot)^2}
                             {(R_{\mathrm{core}}/R_\odot)(L_{\mathrm{core}}/L_\odot)}\,. 
\end{align*}
The nuclear timescale simply measures the time it takes for a certain fraction of
a star's hydrogen to be converted to helium. The numerical factor in 
$\tau_{\mathrm{Nuc}}$ results from ad hoc assuming $10\%$ of the star's 
hydrogen to be converted to helium at the prevailing luminosity.
Once a star derives its energy from an ever thinner nuclear burning shell, 
the concept of the \emph{local }nuclear timescale seems more appropriate 
than the crude $\tau_{\mathrm{Nuc}}$ measure.  
Here, the local nuclear timescale measures the time it
takes to burn away a fraction of the fuel in the nuclear active shell with 
the physical quantities being  evaluated at the maximum 
of the nuclear energy generation rate, $\varepsilon_\mathrm{max}$, of the burning 
shell. Its unsteady temporal runs seen in Fig.~\ref{fig:Timescales} are owed to 
the spatial gridding and adopting $\varepsilon_\mathrm{max}$ at the
respective computational cell, rather than spatially interpolating the maximum.
The abundances are derived at the cell with the maximum nuclear energy generation rate. 
Depending on the motion of this maximum across the computational grid and the
shifting convection boundaries, small jumps are inflicted. 
The $\tau_{\mathrm{Nuc/loc}}$ definition serves as a
lower bound of this timescale because \emph{per definitionem }nuclear burning is assumed
to operate at maximum magnitude in an actually sharply peaked burning shell. 
The global Kelvin-Helmholtz timescale, $\tau_{\mathrm{KH}}$,
measures the time available to support a star's luminosity 
when tapping its internal energy only. 
Restricting the radiating sphere to the inert helium core of 
a shell-burning star leads to the definition of the Kelvin-Helmholtz timescale 
of the core, $\tau_{\mathrm{KHc}}$. The e-folding times of the unstable 
monotonic secular mode(s) (Sec) are displayed as dark-red dots. Finally, the  
continuous blue $\tau_{\teff}$ curve (Te) shows how the transverse 
effective-temperature evolution of the model stars compares to the 
other timescales.

In both panels of Fig.~\ref{fig:Timescales}, the blue $\tau_{\teff}$ (Te) 
loci of the post~--~main-sequence phases
of the $2.5\,\msol$ sequence on the left and the $7\,\msol$ track on the right
are generously sandwiched between $\tau_{\mathrm{Nuc}}$ (Nuc) and 
$\tau_{\mathrm{KH}}$ (KH).
Evidently, $\tau_{\mathrm{Nuc}}$ is much too long and $\tau_{\mathrm{KH}}$ is
significantly too short. The red $\tau_{\mathrm{Nuc/loc}}$ (Nuc/loc) lines on the 
other hand agree better with the stars' evolution across the Hertzsprung gap. 
This applies particularly well to the more massive example where a significant 
speed-up is observed for the model stars between the end of the S-bend and the FGB.
The worsening agreement of the temporal evolution of $\tau_{\teff}$ and 
$\tau_{\mathrm{Nuc/loc}}$ during the $2.5\,\msol$ model's approach 
of the base of the FGB mostly is partially attributable to the luminosity 
drop (Fig.~\ref{fi:AllTracks}), which is not accounted for in the 
$\tau_{\teff}$ (Fig.~\ref{fig:Cmpre_Speed}) definition. 
The timescale behaviors of the $7\,\msol$ example of the right panel of 
Fig.~\ref{fig:Timescales} are representative of all evolutionary 
sequences with significant speed-up after their main-sequence phase.

The dark red points (Sec) added to Fig.~\ref{fig:Timescales} trace the e-folding
times of the unstable monotonic secular mode customary associated with the
SC-instability. The blue edge of the linear instability region lies at 
the divergence to infinity of the e-folding time. 
Red edges were not encountered in any of the analyzed model sequences. 
As to be expected, the secular
e-folding times are much shorter than the nuclear timescale but they
are also considerably longer than $\tau_{\mathrm{KH}}$ the Kelvin-Helmholtz 
timescale of the complete star. In the case of the $2.5\,\msol$ sequence, 
the e-folding time of the secular instability and $\tau_{\mathrm{KH}}$
become comparable only around the base of the 
FGB.\sidenote[][-1.8cm]{The enhanced instability is induced by an avoided crossing
of the originally unstable secular mode with a mode from a subspectrum that 
develops as the model sequence evolves through the base of the FGB. 
See Fig.~\ref{fig:secularModaldiagrm}.}
Hence, most of the Hertzsprung gap is crossed faster than what the unstable 
secular mode supports. 
On the other hand, in the post~--~main-sequence evolution of the
$7\,\msol$ star, the e-folding time of the unstable secular mode is comparable
with $\tau_{\teff}$ on the hot side of the Hertzsprung gap. However, during the
crossing of the gap, the timescale of the secular mode remains longer than
that of the transverse evolution. Even a comparison with 
$\tau_{\mathrm{KHc}}$ (KHc), the Kelvin-Helmholtz timescale of the 
inert stellar core does not help. In the case of the $2.5\,\msol$ 
evolution $\tau_{\mathrm{KHc}}$ is mostly shorter than $\tau_{\mathrm{KH}}$.
Even if $\tau_{\mathrm{KHc}}$ exceeds $\tau_{\mathrm{KH}}$ and approaches 
$\tau_{\teff}$, its temporal functional variation does not match that of 
$\tau_{\teff}$. Therefore, there is no
obvious reason to conjecture that thermal behavior of the inert core drives
the rapid evolution across the HR plane. 

The most important aspect here is that neither the e-folding time 
of the unstable secular mode nor the Kelvin-Helmholtz timescale of 
neither the complete star nor the core alone do correlate with the 
$\tau_{\teff}$ evolution. In particular, the model stars evolve 
fast already \emph{before }any secular instability unfolds (in the QHE case)
or even without any secular instability being possible (in the FE case).

An FE model can be thought of consisting of an inert, isothermal 
core of specified mass and a radiative-zero envelope 
(at least for sufficiently hot model
stars).\sidenote[][-5cm]{Except for the partial ionization regions close to the 
stellar surface, the FE and also the QHE envelopes both admit 
$\diff \log T / \diff \log P \approx 0.23$,
which is close to a radiative-zero solution. Early on, during the 
post~--~S-bend evolution the stars are hot enough so that the partial 
ionization regions and the accompanying convection zones are very 
thin and therefore without significant effect on the global envelope structure. 
} 
The two regions are separated by a nuclear-burning shell. 
The close proximity of evolutionary tracks of FE and QHE models across
the Hertzsprung gap means that both respective envelopes transport roughly 
same luminosity.  In QHE models, 
\emph{nuclear }energy generation produces a higher luminosity than in 
FE models at the same location on the HR plane. Some of energy goes into
the expansion of the envelope so that the surface luminosity of the FE and
QHE models are eventually roughly the same (as it is reflected in the very similar 
computed envelope structures). Hence, the reason of the different timescales of
their evolution through the Hertzsprung gap must be sought in the properties
of the nuclear-burning shell and possibly their effect on the stellar radius. 

\begin{marginfigure} 
	\begin{center}
	\includegraphics[width=0.9\textwidth]{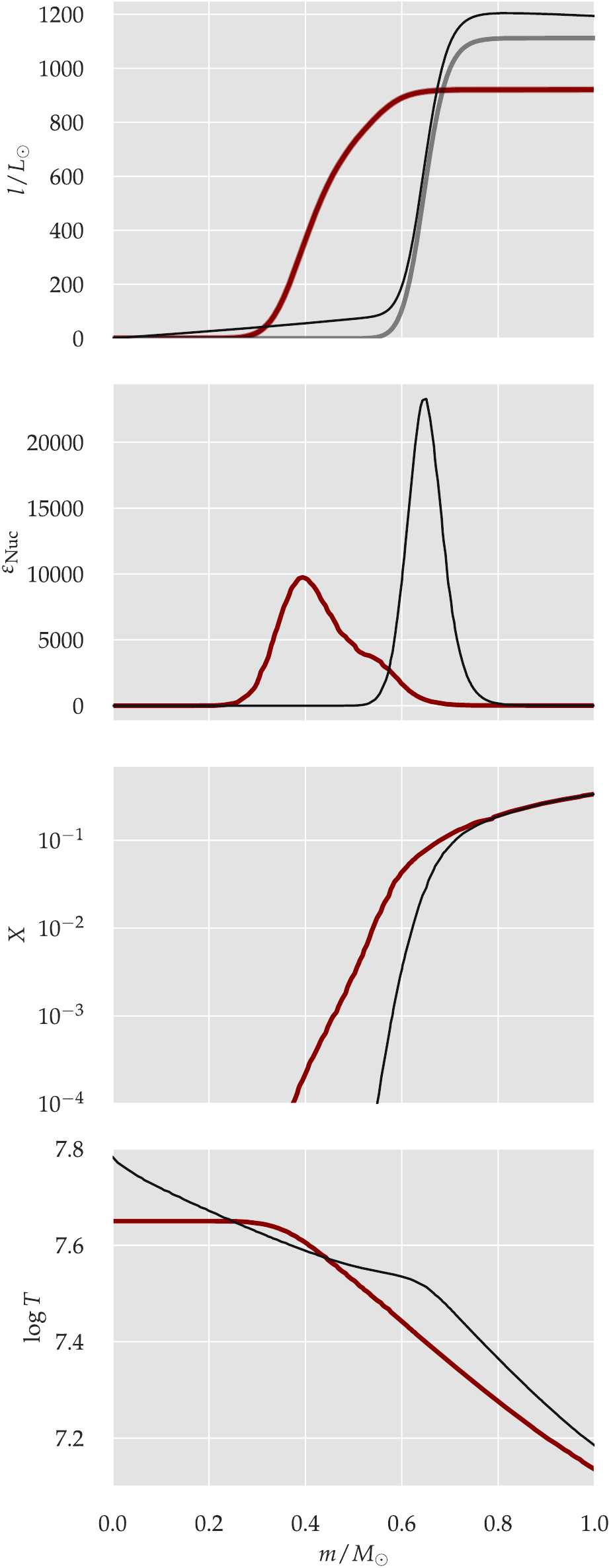} 
	\end{center}
     \caption{Structural difference at around the H-burning shell of $5\,\msol$ 
              QHE (black) and FE (red) models at $\log\teff = 3.8$.
             } \label{fig:M050Z02_QHE_FE}
\end{marginfigure}

In FE models, the nuclear-burning shell alone must adapt in such a way 
as to supply the required transportable energy flux at the base of the envelope.
Because of the developing isothermal core (bottom panel 
in Fig.~\ref{fig:M050Z02_QHE_FE}) and the steeper temperature gradient 
on top of it most of the luminosity is generated at the very inner 
edge of $X$-abundance ramp
(cf.second panel from the bottom in Fig.~\ref{fig:M050Z02_QHE_FE} where 
$X < {\cal O}(10^{-3})$). Apparently, the nuclear shell source of FE models 
is not as strong as that in comparable QHE ones but it is more massive so that
eventually about the same surface luminosity results. The top panel of
Fig.~\ref{fig:M050Z02_QHE_FE} shows the FE-luminosity in red, the nuclear 
luminosity of the QHE model in grey and the total luminosity as a fine black line.
The latter contains the contribution from the star's contraction/expansion, 
which is well expressed around the H-burning shell. At the surface of 
the model star displayed in Fig.~\ref{fig:M050Z02_QHE_FE}, the thin black line 
drops even slightly below the red one. 
Because $X \vert_{\varepsilon_{\mathrm{max}}}$ 
is much smaller in FE models than in QHE ones, their lateral 
timescale ($\tau_{\mathrm{Nuc/loc}}\sim X / \varepsilon_{\mathrm{Nuc}}$)
is accordingly shorter. This means that FE models evolve even faster 
than QHE ones.
  
Early into the post-ms evolution, the mass of the H-burning shell increases 
with stellar mass. The difference in mass diminishes as the stars 
approach the base of the FGB. Hence, 
\emph{the thinner the nuclear burning shell the shorter the local nuclear timescale, 
	  the faster the star's evolution between the end of hydrogen core-burning 
	  and the FGB. } 
Therefore, evolution to the base of the giant branch of the lower-mass stars 
is comparatively slower. 

\bigskip

Even though the usefulness of linear secular stability analyses is limited or even
questionable during fast evolutionary phases we resorted to them here. 
The Appendix at the end of this exposition is devoted to a presentation of the 
mode diagrams encountered in secular eigenmode analyses and
more importantly to tests performed
with \mesa~illustrating how well (or badly) a linear stability analysis does
during the speedy evolution through the Hertzsprung gap.

\section*{Summary}
The evolution of intermediate-mass and massive stars speeds up significantly
after the S-bend at the end of the main-sequence phase and the base of the FGB. 
The resulting scarcity of stars on the HR plane during this evolutionary phase 
is known as the Hertzsprung gap. We showed that also stars in 
\emph{full equilibrium} speed up on the HR plane once a H-burning shell develops. 
Even though a water-tight proof is still pending the results suggest that the
development of the Hertzsprung gap does not depend on a thermal/secular instability. 
The fast evolution across the HR plane can be attributed to the 
properties of the ever thinner H-burning shell at the base 
of the stars' envelopes alone. Hence, the SC instability is rather a secular 
instability that develops \emph{on top }of independently ever faster evolving stars. 
Because \mesa~solves the equations of stellar structure and 
chemical composition simultaneously it is able to track FE models longer 
than the older generations of stellar evolution codes 
that decoupled stellar structure and nuclear evolution. 
The comparison of the QHE and the FE tracks of intermediate-mass model 
stars computed with \mesa~show that they remain close to each other. 
Hence, once it develops, the SC secular instability does not significantly 
affect the stars' evolutionary locus on the HR plane. Even though the 
stellar luminosity profiles are locally affected by the 
secular instabilities, the stars' loci on the HR plane are very 
similar for QHE and FE models, respectively.  
The FE models (where they exist in the Hertzsprung gap) evolve even 
faster than the QHE siblings. Hence the speed-up cannot be attributed 
to a thermal instability of post~--~main-sequence stars.

\newthought{Acknowledgments}: Access to NASA's Astrophysics Data System was
indispensable to write this exposition. Figure 1 relies on data from 
the European Space Agency (ESA) mission {\it Gaia}, 
which is processed by the {\it Gaia} Data Processing and Analysis 
Consortium (DPAC).
Funding for the DPAC has been provided by national institutions, 
in particular the institutions participating in the {\it Gaia} 
Multilateral Agreement. The python code listed in a 2018 blog post 
of Vlas Sokolov (\texttt{https://vlas.dev/post/}) served as a 
helpful template to efficiently start querying the 
{\it Gaia} DR2 database to produce Fig.~\ref{fig:GaiaCMD}. 
The model stars were computed with the Modules for Experiments in 
Stellar Astrophysics 
\citep[\mesa, ][and references therein]{Paxton2018} mostly in version $12\,115$. 

\medskip

\noindent This installment is dedicated to the memory of P.~Freiburghaus, 
aka H.H.~--~he was an invaluable source of inspiration. 

\section{Appendix}
Secular stability of the model stars of the various evolutionary tracks 
was computed in the linear regime with the Riccati approach as described 
in \citet{alfalt07}. Again, the QHE approximation was used, 
i.e. a simple perturbation of the $\diff_t Q$ term was included in
the secular stability problem; convection was treated as instantaneously 
adapted to its environment. 

\newthought{The structure of the secular mode diagrams}
of intermediate-mass and massive stars during 
hydrogen core- and shell-burning phase, representatively illustrated 
by the  $2.5 \msol$ 
star.\sidenote[][-1.8cm]{The temporal separation ansatz adopts $q(m;t) 
			  \propto \exp(\imag \sigma  t)$ with $\sigma$ being
			  possibly complex $(\sigr,\sigi)$ because the secular problem
			  is not self-adjoint. The eigenfrequencies $\sigma$ are expressed
			  in units of the star's free-fall frequency.}

\begin{figure} 
	\begin{center}
	\includegraphics[width=0.90\textwidth]{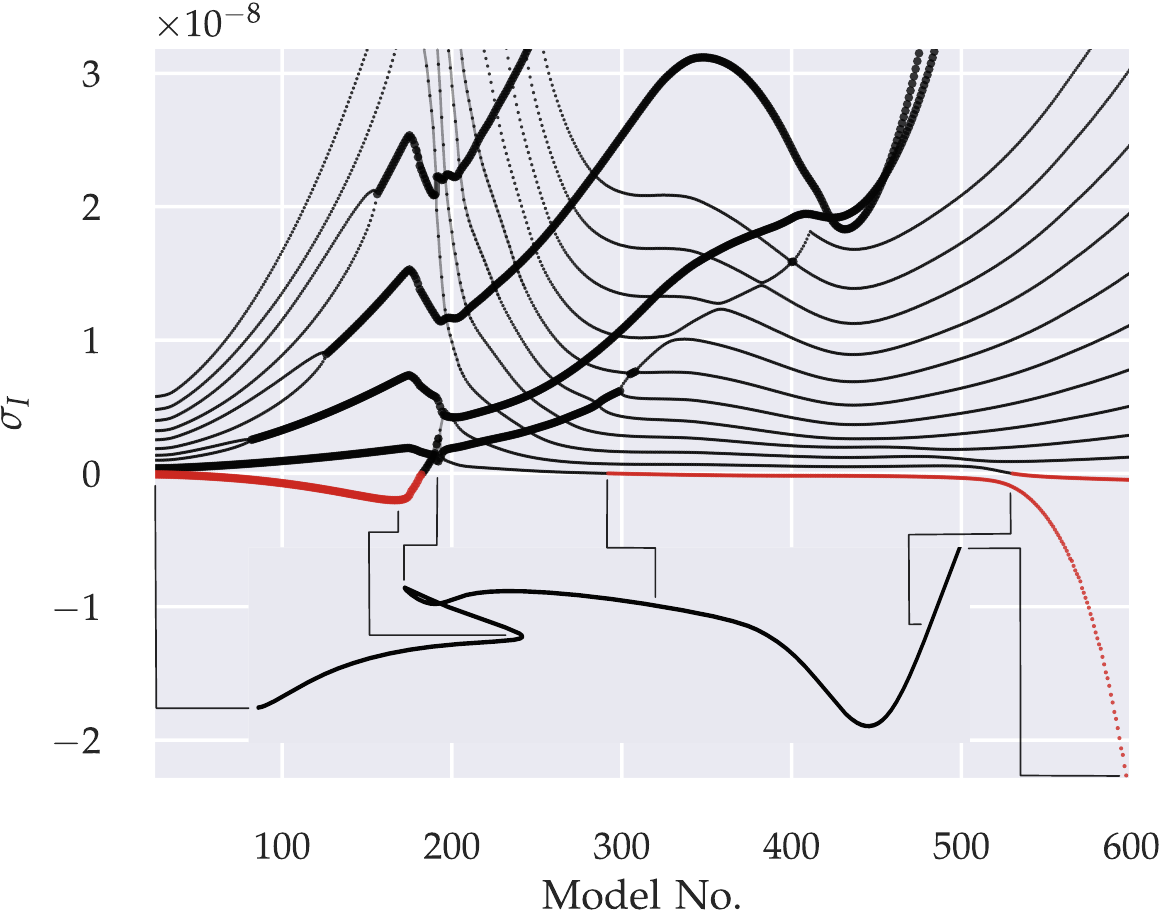} 
	\end{center}
     \caption{Secular mode diagram, imaginary parts of the eigenvalues
     		  as a function of model number along the evolutionary track of
     		  the $2.5\,\msol$ star inset at the bottom of the figure.
     		  Small circles mark monotonic eigenmodes
     		  ($\sigma = \left(0,\sigma_\mathrm{I}\right)$); the stronger
     		  filled circles indicate complex eigensolutions.  
     		  Black colored eigenvalues mark damped modes, red ones unstable
     		  ones ($\sigma_\mathrm{I} < 0$).
     		  For better orientation along the mode-number abscissa, a 
     		  few pertinent epochs along the HR-track of the star 
     		  point to the respective positions in the secular mode-diagram.
     		  Except for the avoided crossing in the lower right the mode 
     		  diagram is representative for all evolutionary sequences
     		  computed for this exposition.   
             } \label{fig:secularModaldiagrm}
\end{figure}

Already during the main-sequence phase of the star, the lowest two eigenmodes are 
complex, i.e. the two secular modes are oscillatory. The lowest mode is even 
weakly unstable. Except for the $10\,\msol$ sequence,
this behavior has been found for all masses considered here. 
Despite reducing the evolutionary timesteps of \mesa~models well below the 
stars' Kelvin-Helmholtz timescale during respective main-sequence phases, 
no evidence of an instability could be found in the evolution computations. 

As evolution progresses along the main sequence, pairs of ever 
higher-order monotonic eigenmodes merge into complex eigensolutions 
as it has been reported in the literature for a 
long time \citep[e.g.][]{Aizenman1971a}. 
Once a star approaches the FGB, all $\sigi$ tend to grow again. 
It is only the lowest monotonic secular eigenmode, i.e. the lower branch 
of the unfolded, initially complex eigenmode that goes unstable.
In the case of the $2.5\,\msol$ sequence actually two secular modes are 
found unstable at the lower FGB. The mode of a secular subspectrum that develops
during the shell-burning phase of evolution
interacts with the mode family running almost horizontally 
in Fig.~\ref{fig:secularModaldiagrm}. At the avoided crossing with the lowest-order,
already unstable secular mode the interfering mode picks up a large negative  
$\sigi$ and develops into the dominantly unstable mode.

The way the lowest monotonic mode branch goes unstable is 
similar in all computed mass sequences:
During the main-sequence phase, the lowest eigenmodes are complex, 
either already on the ZAMS or they merge into pairs during core hydrogen burning. 
The complex modes react only weakly on the restructuring of the star as 
nuclear burning shifts from center to a shell. When the 
nuclear-burning shell develops, an additional sub-spectrum of purely monotonic 
eigenmodes intrudes from large $\sigi$ values to form a broad trough 
as the star evolves across the Hertzsprung gap. Also around the onset 
of shell burning, the lowest complex mode unfolds into a pair of 
monotonic modes with the lower-$\sigi$ branch going unstable during 
the post S-bend evolution. It is this monotonic instability that is 
commonly associated with the SC-instability.  

\begin{marginfigure} 
	\begin{center}
	\includegraphics[width=1.0\textwidth]{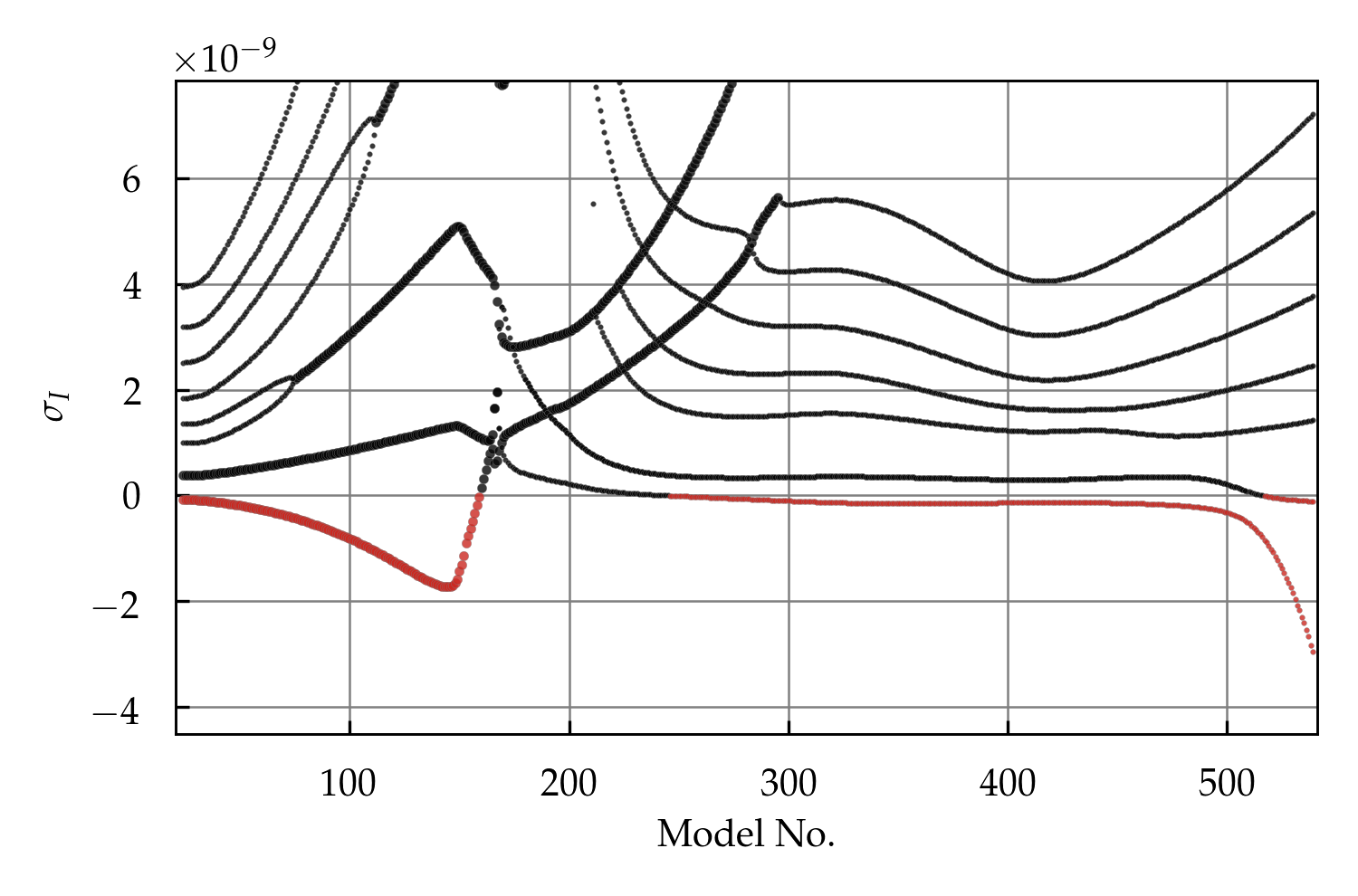} 
	\end{center}
     \caption{Fragmentary secular mode diagram for the $2.5\,\msol$ 
     		  model track computed with the Schwarzschild criterion and 
     		  without overshooting. N.B. the model numbers in the plot 
     		  belong to different evolutionary epochs from those appearing 
     		  in Fig.~\ref{fig:secularModaldiagrm}. 
             } \label{fig:M025Z02_SSD}
\end{marginfigure}
\citet{gabriel72} contemplated that the differences in the 
secular eigenspectrum of his study and the work of \citet{Aizenman1971} 
might be attributable to differences in the central convection zones. 
Our \emph{standard }setup in \mesa~relied on the Ledoux criterion with moderate
overshooting at the convective boundaries. 
To scrutinize the robustness of \emph{our }secular results
with respect to convection treatment 
we recomputed the $2.5 \msol$ track adopting the Schwarzschild 
criterion and neglecting convective overshooting. 
The resulting secular mode diagram is shown in Fig.~\ref{fig:M025Z02_SSD}. 
Even though the size and the compositional boundary of the convective stellar 
cores differ between the two model series these changes do not alter
the secular eigenspectrum qualitatively. In the Schwarzschild case, the blue edge of 
the monotonic secular instability shifts to slightly higher effective
temperatures ($\log \teff = 3.87$) but remains at the same relative core
mass of $q=0.09$, which is comparable to magnitude along the Ledoux-treatment 
track. Except for this shift, the structure of the mode diagram remains unchanged. 
As a pertinent side observation, notice that the unstable oscillatory secular mode 
during the main-sequence phase persists. 

\newpage

\newthought{Scrutinizing the Riccati-calculated Instabilities} 
with \mesa: At epochs around 
the emergence of monotonic secular instabilities as computed with the Riccati code, 
\mesa~was restarted enforcing $\diff_t X_i = 0$ for all accounted nuclear species  
$X_i$ but maintaining normal nuclear energy generation otherwise; 
this is referred to as the \emph{no~--~nuclear-evolution }(NNE) approach. 
If an arbitrary QHE model is forced to evolve in the NNE approximation it 
either settles quickly
close to its starting position on the HR plane and the timestep grows without 
bound once the model reaches full thermal equilibrium. In contrast, if the initial QHE 
model is thermally unstable it will evolve, with its frozen compositional
stratification, on a thermal timescale towards its thermal equilibrium structure, 
which usually is not close to its QHE position in the HR diagram. 
\citep[see also][]{Scuflaire1995}.  

\begin{figure*} 
	\begin{center}
	\includegraphics[width=0.98\textwidth]{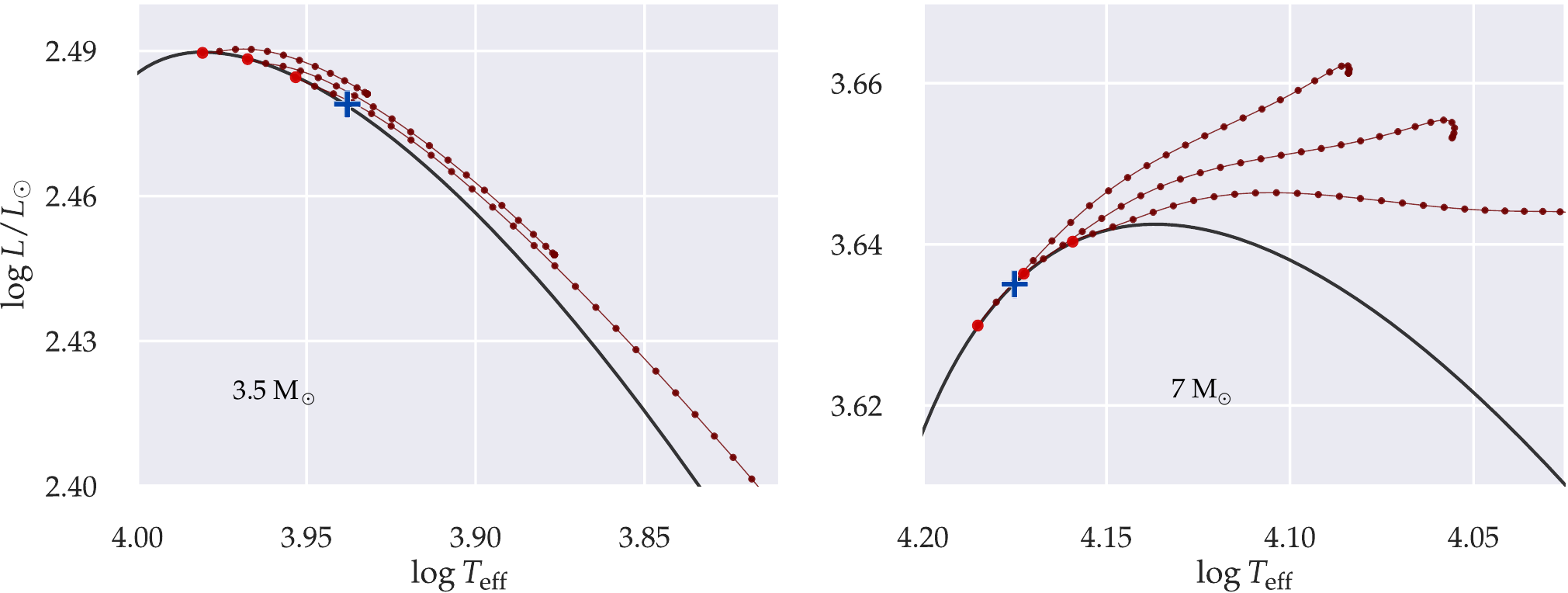} 
	\end{center}
     \caption{Two examples of comparing secular stability as evidenced in 
     \mesa~directly as compared with the appearance of an unstable monotonic mode in 
     the linear secular approximation.
             } \label{fig:RiccatiValidation}
\end{figure*}

The NNE ansatz was invoked at a few selected epochs~--~in the neighborhood
of the blue edge of the monotonic secular-mode instability as computed with
the Riccati code (shown as blue crosses in 
Fig.~\ref{fig:RiccatiValidation})~--~for the 3.5 and the 7 $\msol$ models, 
respectively. The starting epochs are marked by red dots on the QHE 
evolutionary track in Fig.~\ref{fig:RiccatiValidation}. 
A few models along the NNE tracks 
are plotted as connected smaller dark-red dots. If the starting model is 
thermally stable, the corresponding NNE track does not significantly drift 
off the QHE track but quite quickly stalls and the timestep of the computation 
grows without bound (because nothing happens anymore to the star). 
If the initial model is thermally unstable, though, its
NNE evolution takes it to a new thermally stable configuration which can be far
from the initial QHE location on the HR plane. In both panels of 
Fig.~\ref{fig:RiccatiValidation}, the earlier two starting models along the
3.5 and the 7 $\msol$ sequences are both thermally stable.
The third NNE loci of both sequences diverge  
from the respective QHE tracks and head towards the FGB far outside the figures. 
Even though the linear secular analyses do not pin down the onset of the
instability very accurately, they remain nonetheless helpfully indicative. 

Along the FGB, the situation is less obvious:
As mentioned before, the linear secular stability analyses failed to find red edges
in any of the computed model sequences. Is this real
or a problem of the Riccati code? Once again we ran the NNE experiment, 
this time along the lower FGB branches. 
\begin{figure*} 
	\begin{center}
	\includegraphics[width=0.98\textwidth]{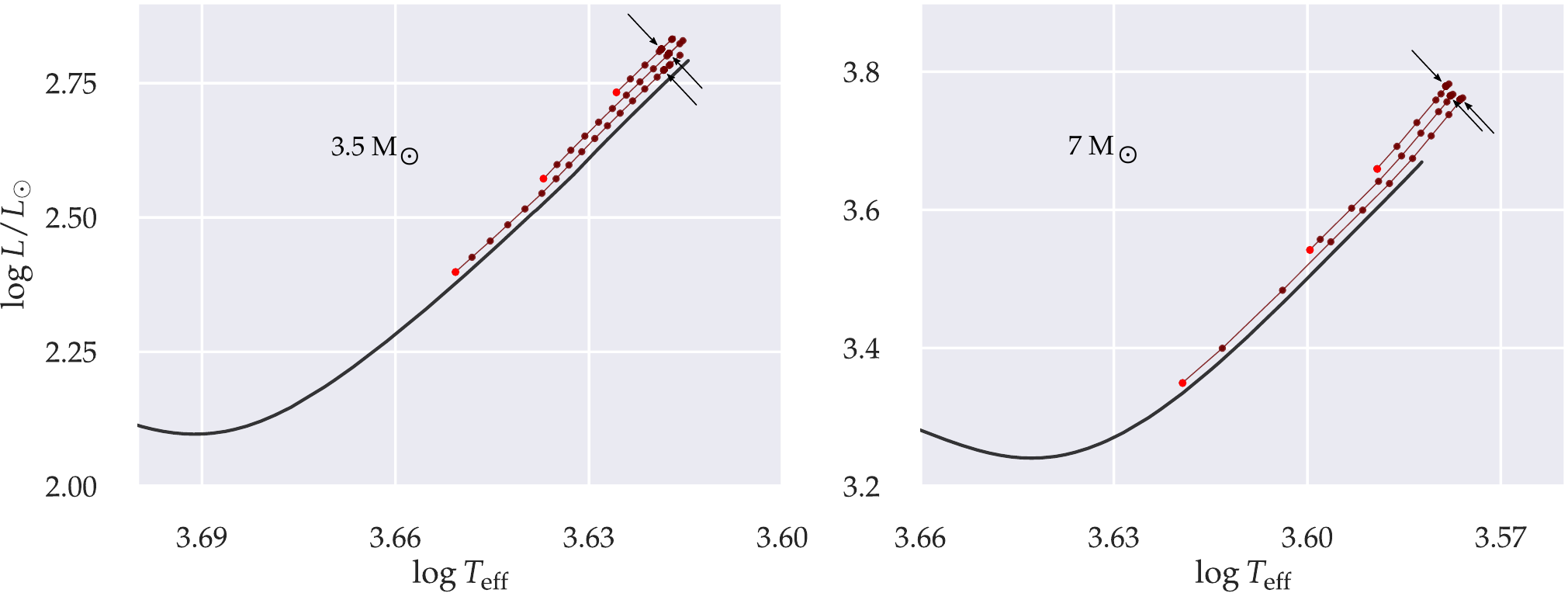} 
	\end{center}
     \caption{Comparable to Fig.~\ref{fig:RiccatiValidation} but now in the region
     		  of the FGB where the red edge of the unstable monotonic 
   			  secular mode was expected not found. 
   			  For clarity, the NNE tracks along the FGB are each shifted by 
   			  multiples of $0.002$ in $\log \teff$. 
   			  Without shift all tracks fall essentially onto the QHE FGB locus.
   			  Arrows point to the positions of the terminal locations where the
   			  NNE tracks stalled.    			  
             } \label{fig:RiccatiValidation-FGB}
\end{figure*}

Figure~\ref{fig:RiccatiValidation-FGB} zooms in to the lower FGBs of the 3.5 (left 
panel) and the 7 $\msol$ (right panel) QHE tracks (shown as black lines). The tracks
stop at ignition of core He-burning. Superposed on the QHE tracks are, as
in Fig.~\ref{fig:RiccatiValidation}, NNE tracks that started at three different epochs
(bright red dots) along the FGB. In both panels of
Fig.~\ref{fig:RiccatiValidation-FGB} the NNE tracks are successively 
shifted in $\teff$ to avoid overlapping. All NNE tracks followed very closely 
the locus of the QHE one. In all cases the NNE models drifted away from their 
initial QHE position on the HR plane. On a thermal timescale they all evolved up 
the FGB and all stalled at about the same luminosity. In the 3.5~$\msol$ case the
terminal luminosity as close to that of the QHE epoch where core He-burning ignites.
Inspection of central  $\rho, T$ of the final NNE models reveals 
that they too would be ready to start core He-burning. The very 
definition of NNE though prevents them from changing their nuclear composition.
Therefore, NNE evolution \emph{must }stall along the FGB. The same applies to the
7~$\msol$ sequence even though in NNE models He-burning ignition conditions are 
only reached at luminosities that are somewhat higher than those of the QHE models. 
 
If stellar models are thermally unstable~--~in the NNE sense~--~depends on the 
definition of a suitable neighborhood that is to be left upon starting NNE evolution 
at a chosen QHE epoch. Along the FGB such a neighborhood is 
particularly ill-defined because it seems constrained by the evolution 
along the Hayashi track and the suppression of core
He-burning. Nonetheless, at the very least the NNE models from the earliest two
epochs around the base of the FGB evolve away considerably from their QHE location
(in both stellar-mass examples) so that they likely qualify as secularly unstable. 
If the same applies to the respective last QHE epochs must remain open for the moment. 

\bibliographystyle{aa}
\bibliography{Hertzsprung}        

\end{document}